\documentclass{ws-procs975x65}

\begin{document}

\title{\Large Self-Dual Fields on the space of a Kerr-Taub-bolt Instanton}

\author{ALIKRAM  N. ALIEV\footnote{aliev@gursey.gov.tr}}

\address{Feza G\"ursey Institute, P.K. 6  \c Cengelk\" oy, 34684 Istanbul, Turkey}
\author{ CIHAN SA\c{C}LIO\u{G}LU\footnote{saclioglu@sabanciuniv.edu}}
\address{Faculty of Engineering and Natural Sciences, Sabanci
University, Tuzla, 81474 Istanbul, Turkey}

\begin{abstract}
We discuss  a new exact solution for self-dual Abelian gauge
fields living on the space of the Kerr-Taub-bolt instanton, which
is a generalized example of asymptotically flat instantons with
non-self-dual curvature.
\end{abstract}
\bodymatter
\section{Introduction}
Gravitational instantons are  complete nonsingular solutions of
the  vacuum Einstein field equations in Euclidean space. The first
examples  of gravitational instanton metrics were obtained by
complexifying the  Schwarzschild, Kerr and Taub-NUT spacetimes
through analytically continuing them to the Euclidean sector
\cite{h,gh}. The Euclidean Schwarzschild and Euclidean Kerr
solutions do not have self-dual curvature though they are
asymptotically flat at spatial infinity and periodic in imaginary
time, while the Taub-NUT instanton is self-dual. There also exist
Taub-NUT type instanton metrics\cite{page, gperry} which are not
self-dual and possess an event horizon ("bolt"). Other examples of
gravitational instanton solutions are given by the multi-centre
metrics \cite{gh}. These metrics are asymptotically locally
Euclidean with self-dual curvature and admit a hyper-K\"ahler
structure. The hyper-K\"ahler structure becomes  most transparent
within the Newman-Penrose formalism for Euclidean
signature\cite{an}.

One of the striking properties of manifolds with  Euclidean
signature is that they can harbor self-dual gauge fields. In other
words, solutions of Einstein's equations automatically satisfy the
system of coupled Einstein-Maxwell and Einstein-Yang-Mills
equations. The corresponding solutions for some  gauge fields and
spinors that are  inherent in the given instanton metric were
ontained in papers \cite{HawkPope, cih}. In recent years, there
has been some renewed interest in self-dual gauge fields living on
well-known Euclidean-signature manifolds. For instance, the gauge
fields were studied by constructing a self-dual square integrable
harmonic form on a given hyper-K\"ahler space\cite{gibbons}. The
similar square integrable harmonic form on spaces with
non-self-dual metrics was found only for the the
Euclidean-Schwarzschild instanton\cite{etesi}. As a generalized
example of this, we shall consider the Kerr-Taub-bolt instanton
and construct a square integrable harmonic form  to describe
self-dual Abelian gauge fields harbored by the instanton.
\section{The Kerr-Taub-bolt instanton}
The Kerr-Taub-bolt instanton\cite{gperry} is a
Ricci-flat metric with asymptotically flat behaviour. It has the
form
\begin{equation}
ds^{2}=\Xi \left( \frac{dr^{2}}{\Delta }+d\theta ^{2}\right)
+\frac{\sin
^{2}\theta }{\Xi }\left( \alpha \,dt+P_{r}\,d\varphi \right) ^{2}+\frac{\Delta }{%
\Xi }\left( dt+P_{\theta }\,d\varphi \right) ^{2}\,\,, \label{ktb}
\end{equation}
where the metric functions are given by
\begin{eqnarray}
\Delta &=& r^{2}-2Mr-\alpha ^{2}+N^{2}\,,~~~~\Xi =  P_{r}-\alpha
P_{\theta }=r^{2}-(N+\alpha \cos \theta
)^{2}\,\,, \nonumber \\[2mm]
P_{r}&=&r^{2}-\alpha ^{2}-\frac{N^{4}}{N^{2}-\alpha ^{2}}\,,~~~~
P_{\theta } =  -\alpha \sin ^{2}\theta +2N\cos \theta
-\frac{\alpha N^{2}}{N^{2}-\alpha ^{2}}\,\,\,.
\end{eqnarray}
The parameters $\, M ,\,\, N ,\,\, \alpha \,\,$ represent the
"electric" mass, "magnetic" mass and "rotation" of the instanton,
respectively. The coordinate $\,t\,$ in the metric behaves like an
angular variable and  in order to have a complete nonsingular
manifold at values of $\,r\,$ defined by equation $\,\Delta=0\,$ ,
$\,t\,$ must have a period $\,2\pi/\kappa\,$. The coordinate
$\,\varphi\,$ must also be periodic with period $\,2\pi
\,(1-\Omega /\kappa )\, $, where the "surface gravity" $ \kappa =
(r_{+}-r_{-})/2\,r_{0}^{2} \,\,$,  the "angular velocity" of
rotation $ \Omega =\alpha/r_{0}^{2} $  and
\begin{eqnarray}
r_{\pm }&=& M \pm \sqrt{M^{2}-N^{2}+\alpha ^{2}}\,\,\,,~~~~~~~
r_{0}^{2} = r_{+}^{2}-\alpha ^{2}-N^{4}/(N^{2}-\alpha ^{2})\,\,\,.
\end{eqnarray}
As a result one finds that the condition $ \kappa =1/(4|N|) $
along with $ \Xi \geq 0 \,$ for $\,r>r_{+}\,$ and $\,0\leq \theta
\leq \pi \,$ guarantees that $\,r=r_{+}\,$  is a regular bolt in
(\ref{ktb}) . Clearly, the isometry properties of the
Kerr-Taub-bolt instanton with respect to a $\, U(1)\,$- action in
imaginary time imply the existence of the Killing vector field $
\partial_{t} = \xi^{\mu}_{(t)}\,\partial_{\mu}\,\,.$
\section{ Harmonic $2$-form}
We shall use the above Killing vector field to construct a square
integrable harmonic $2$-form on the Kerr-Taub-bolt space. We start
with the associated Killing one-form field $ \xi = {\xi}_{(t)
\mu}\,d\,x^\mu $. Taking the exterior derivative of the one-form
in the metric (\ref{ktb}) we have
\begin{eqnarray}
\label{2form}
 d\xi &=&\frac{2}{\Xi^2}\left\{ \,\left[M r^2 +
\left(\alpha M \cos\theta-2 N r+M N\right)\left( N+\alpha
\cos\theta
\right)\right] e^{1} \,\wedge e^{4} \right. \\[2mm]  & & \left. \nonumber
 -\left[N\,\left(\Delta + \alpha^2 +\alpha^2\, \cos^2 \theta\right) + 2\,
\alpha (N^2-M r) \cos\theta\right] e^{2} \,\wedge
e^{3}\,\right\}\,,
\end{eqnarray}
where we have used the basis one-forms satisfying the simple
relations of the Hodge duals:  $\,  ^{\star}\left(e^{1} \,\wedge
e^{4}\right)= e^{2} \,\wedge e^{3}\,,~~  ^{\star}\left(e^{2}
\,\wedge e^{3}\right)= e^{1} \,\wedge e^{4}\,\, $. Straightforward
calculations  show that the two-form (\ref{2form}) is both closed
and co-closed, that is, it is a harmonic form. However the
Kerr-Taub-bolt instanton does not admit hyper-K\"ahler structure,
and the two-form  is not self-dual.
\section{Stowaway fields}
To describe the Abelian "stowaway" gauge fields we define the
(anti)self-dual two form
\begin{equation}
F=\frac{\lambda}{2}\,(d\xi \pm \,^{\star} d\xi)\,, \label{sdual}
\end{equation}
where $\,\lambda\,$ is an arbitrary constant related to the dyon
charges carried by the fields. Using in this expression the
two-form (\ref{2form}) and its Hodge dual we obtain the harmonic
self-dual two-form\cite{ac}
\begin{equation}
F=\frac{\lambda (M-N)}{\Xi^2}\,\left(r+N +\alpha
\cos\theta\right)^2 \left( e^{1} \,\wedge e^{4} + e^{2} \,\wedge
e^{3}\right)\,\,, \label{sd2form}
\end{equation}
which implies the existence of the potential one-form
\begin{equation}
A=- \lambda\,(M-N)\,\left[\cos\theta\, d\varphi + \frac{r+N
+\alpha \cos\theta}{\Xi} \,\left(d t+ P_{\theta}\, d
\varphi)\right)\right] \,\,. \label{spotform}
\end{equation}
From equation (\ref{sdual}) one can also find the corresponding
anti-self-dual two-form. The square integrability of these
harmonic two-forms can be shown by explicitly integrating the
Maxwell action. For the self-dual two-form we have
\begin{equation}
\frac{1}{4\pi ^{2}}\int F\wedge F =\frac{2\lambda ^{2}%
}{\kappa }\,(M-N)\,\left( 1-\frac{2 \,\alpha }{r_{+}-r_{-}
}\right)\,\,. \label{maxact}
\end{equation}
Since this integral is finite, the self-dual two-form $\,F\,$ is
square integrable. The total magnetic flux
\begin{equation}
\Phi=\frac{1}{2\pi }\int_{\Sigma }F = 2\lambda \,(M-N)\,\left( 1-
\frac{2 \,\alpha }{r_{+}-r_{-} }\right)\,,
\end{equation}
must be equal to an integer $\,n\,$ because of  the Dirac
quantization condition. We see that the periodicity of angular
coordinate in the Kerr-Taub-bolt metric affects the
magnetic-charge quantization rule in a non-linear way. It involves
both  the "electric" and "magnetic"  masses and the  "rotation"
parameter.

\vfill
\end{document}